\documentclass[a4,10pt]{article}
\usepackage{amsmath,amssymb}
\usepackage{graphicx}
\begin{document}
\title{ The sum rules for the spin dependent structure functions
corresponding to the moment at $n=0$} 
\author{Susumu Koretune\\
Department of Physics, Shimane University \\ 
Matsue,Shimane,690-8504,Japan}
\maketitle
\begin{abstract}
Sum rules for the spin dependent structure functions corresponding
to the moment at $n=0$ derived from the current algebra based on
the canonical quantization on the null-plane are reviewed. 
\end{abstract}
\section{Introduction}
Many years ago, it was shown that the $O(4)$ partial waves of the $t$ channel CM scattering 
amplitude of the current-hadron reaction at $t=0$
exactly correspond to the Nachtmann moments.\cite{kore1} This fact
makes clear that the Nachtmann moment\cite{Nacht} can be used even
in the small $Q^2$ region where the operator product expansion
can not be applied.
Now the Nachtmann moments at $n=0$ for the spin dependent structure functions 
take a very simple form and its form is the same
as the one which appears in the fixed-mass sum rules from the current algebra based on the
canonical quantization on the null-plane. Thus we can treat these sum
rules without worrying about the kinematical target mass correction. 
The sum rules for the spin dependent structure functions $g_1$ and $g_2$ 
in the small $Q^2$ region derived in Refs.\cite{kore2,kore3,kore4} 
belong to this category. These sum rules show that there is a
connection among the resonances, the elastic, and the nonresonant continuum 
in the $g_1$ and the $g_2$ independently. 
Since the Born term changes rapidly in the small $Q^2$ region,
the sum of the resonance and the nonresonant continuum also changes rapidly.
In case of the $g_1$, this explains why the Gerasimov-Drell-Hearn(GDH) sum changes a sign
in the very small $Q^2$ region. In case of the $g_2$, we find that the Born term divided by
$Q^2/2$ in the sum rule has a very similar behavior with that in the sum rule for the
$g_1$.  This divided Born term is nothing but the Born term in the Schwinger sum
rule for the $g_2$,\cite{Schw} which was derived also by the fixed-mass sum rule
approach.\cite{DJT} These analysis show that, when we consider the duality like
Bloom-Gilman,\cite{BG} the proper inclusion of the elastic contribution
is indispensable.  In this paper, we give a brief review of these sum rules.

\section{Sum rules from the current algebra based on the canonical quantization on the null-plane}
The spin dependent part of the hadronic tensor of the imaginary part of 
the forward current-hadron scattering amplitude is defined as 
\begin{eqnarray}
W^{\mu\nu}_{ab}|_{spin}&=&i\epsilon^{\mu\nu\lambda\sigma}q_{\lambda}s_{\sigma}G_1^{ab}
+i\epsilon^{\mu\nu\lambda\sigma}q_{\lambda}(\nu s_{\sigma}-
q\cdot sp_{\sigma})G_2^{ab}\\\nonumber
&=&\frac{1}{4\pi}\int d^4x\exp (iqx)
<p,s|[J^{\mu}_a(x),  J^{\nu}_b(0)]|p,s>_c |_{spin}.
\end{eqnarray}
The structure function has a crossing symmetry
$G_1^{ab}(p\cdot q,q^2)=-G_1^{ba}(-p\cdot q,q^2)$ and $G_2^{ab}(p\cdot
q,q^2)=G_2^{ba}(-p\cdot q,q^2)$
under $q\to -q$, $a \leftrightarrow b$, and $\mu
\leftrightarrow \nu$.    
The spin dependent part of the fixed-mass sum rule from the current
algebra based on the canonical quantization on the
null-plane \cite{DJT} can be obtained as follows. We take
$x^{\pm}=\frac{1}{\sqrt{2}}(x^0\pm x^3)$.\\
\begin{minipage}{6cm}
\includegraphics[width=5.7cm]{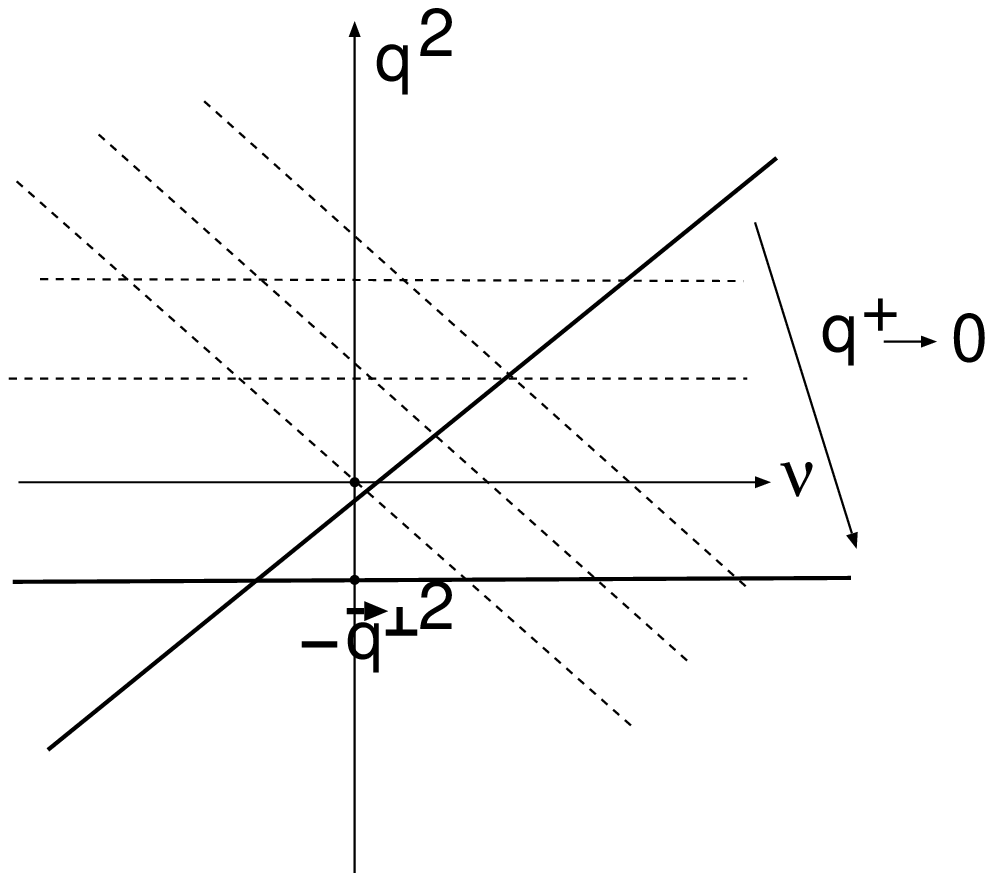}
\end{minipage}
\begin{minipage}{6cm}
We first set $\mu = +$ and $\nu = i$. Then,
on the left hand side of Eq.(1), we integrate over $q^-$ 
and change variable from $q^-$ to
$\nu = p^+q^- + p^-q^+ -\vec{p}^{\bot}\vec{q}^{\bot}$.
Then we obtain
$q^2 = 2q^+\frac{\nu - p^-q^+ +\vec{p}^{\bot}\vec{q}^{\bot}}{p^+}
 -\vec{q}^{\bot 2} .$ 
We set $q^+=0$ before $\nu$ integration. To assure this
manipulation, we need the superconvergence
relation to neglect the contributions from the time-like $q^2$ region. 
On the right hand side, we see that the commutation relation is restricted
at $x^+=0$ by interchanging the $x$ integration and the $q^-$
integration. It is known that this interchange 
\end{minipage}
is possible under the
same superconvergence relation which we need on the left hand
side.\cite{DJT,CJFR}
Now, we take the current formed by the quark bilinear as
$J^{\mu}_{a}(x)=\bar{q}(x)\gamma^{\mu}\frac{\lambda_a}{2}q(x)$.
The quark field on the null-plane
is decomposed by the projection operator as
\begin{equation}
 q^{(\pm )}(x)=\Lambda^{\pm}q(x),\quad \Lambda^{\pm}=\frac{1}{2}(1\pm \gamma^0\gamma^3).
\end{equation}
The $q^{(-)}(x)$ is expressed by the $q^{(+)}(x)$ through the equation of the
motion, and hence only the $q^{(+)}(x)$ is the independent quantity. 
The canonical anti-commutation relation on the null-plane is defined as
\begin{equation}
 \{q^{(+)\dagger}(x),q^{(+)}(0)\}|_{x^+=0}=\sqrt{2}\Lambda^{+}\delta^2(\vec{x}^{\bot})\delta({x^-}).
\end{equation}
The current $J^+_{a}(x)$ is given only by the $q^{(+)}(x)$
\begin{equation}
 J^{+}_a(x)=\bar{q}(x)\gamma^{+}\frac{\lambda_a}{2}q(x)=\sqrt{2}q^{(+)\dagger}(x)\frac{\lambda_a}{2}q^{(+)}(x).
\end{equation}
On the other hand, $J^{i}_a(x)$ is given as
\begin{equation}
J_a^i(x)=q^{(+)\dagger}(x)\gamma^0\gamma^i\frac{\lambda_a}{2}q^{(-)}(x)
+q^{(-)\dagger}(x)\gamma^0\gamma^i\frac{\lambda_a}{2}q^{(+)}(x).
\end{equation}
Then, the following commutation relation holds in QCD.
\begin{eqnarray}
 [J_a^{+}(x),J_b^i(0)]|_{x^+=0}&=&i[s^{+\beta
 i\alpha}\partial_{\alpha}[\Delta(x)G_{c\beta}(x|0)]\\\nonumber
&-&2g^{+\alpha}g^{i\beta}\partial_{\alpha}[\Delta(x)]G_{c\beta}(x|0)\\\nonumber
&-&\epsilon^{+i\alpha\beta}\partial_{\alpha}[\Delta (x)G_{c\beta}^5(x|0)]],
\end{eqnarray}
where
\begin{equation}
 \Delta (x)|_{x^+=0}=-\frac{1}{4}\epsilon
 (x^-)\delta^2(\vec{x}^{\bot}),
\end{equation}
\begin{equation}
G_c^{\beta}(x|0)=d_{abc}A_c^{\beta}(x|0)+f_{abc}S_c^{\beta}(x|0),\quad
G_c^{5\beta}(x|0)=d_{abc}S_c^{5\beta}(x|0)-f_{abc}A_c^{5\beta}(x|0),
\end{equation}
and
\begin{eqnarray}
S_a^{\mu}(x|0)=\frac{1}{2}[\bar{q}(x)\gamma^{\mu}\frac{\lambda_a}{2}q(0)+\bar{q}(0)\gamma^{\mu}\frac{\lambda_a}{2}q(x)],\\\nonumber
A_a^{\mu}(x|0)=\frac{1}{2i}[\bar{q}(x)\gamma^{\mu}\frac{\lambda_a}{2}q(0)-\bar{q}(0)\gamma^{\mu}\frac{\lambda_a}{2}q(x)].
\end{eqnarray}
The sum rule for the spin dependent structure functions $g_1$ and $g_2$ can be derived from
this relation by comparing the coefficient of $p^i$ and $q^i$, where $\nu G_1^{ab}=g_1^{ab}$ and
$\nu^2G_2^{ab}=g_2^{ab}$. In terms of the $g_1$ and the $g_2$, the result
given in Ref.\cite{DJT} can be written as
\begin{equation}
 \int_{0}^{1}\frac{dx}{x}g_1^{[ab]}(x,Q^2)=-\frac{1}{16} f_{abc}\int_{-\infty}^{\infty}
d\alpha \epsilon (\alpha )[A_c^5(\alpha ,0)+\alpha \bar{A}_c^5(\alpha ,0)],
\end{equation}
\begin{equation}
 \int_{0}^{1}\frac{dx}{x}g_2^{[ab]}(x,Q^2)=\frac{1}{16} f_{abc}\int_{-\infty}^{\infty}
d\alpha\epsilon (\alpha ) \alpha \bar{A}_c^5(\alpha ,0),
\end{equation}
\begin{equation}
 \int_{0}^{1}dxg_2^{(ab)}(x,Q^2)=0,
\end{equation}
where
\begin{equation}
 <p,s|A_c^{5\beta}(x|0)|p,s>_c=s^{\mu}A_c^5(p\cdot x,x^2)+p^{\mu}(x\cdot
 s)\bar{A}_c^5(p\cdot x,x^2)+x^{\mu}(x\cdot s)\tilde{A}_c^5(p\cdot
 x,x^2),
\end{equation}
and, for $i=1,2$,
\begin{equation}
g_i^{(ab)}=\frac{1}{2}(g_i^{ab}+g_i^{ba}),\quad
 g_i^{[ab]}=\frac{1}{2i}(g_i^{ab}-g_i^{ba}).
\end{equation}
Corresponding to the sum rule (10), Beg sum rule where the right
hand side of Eq.(10) was zero had been known
in the equal-time formalism and considered to be peculiar since it was invalid in the free
field model. This fact was discussed in
Ref.\cite{Beg}, and also in Ref.\cite{Adler}. The modification which
appears on the right hand side of the sum rule (10) in the null-plane method
circumvented the defect. 
The sum rule (12) in case of the virtual photon was derived by
Schwinger\cite{Schw} and also by Burkhart and Cottingham.\cite{BC} 
At large $Q^2$, since the Born term can be
neglected, the sum rule (12) holds for the inelastic reaction.
and it has been verified by the perturbative QCD.\cite{alta,kodaira,KT}
\section{Sum rules for the $g_1$ and the $g_2$ in the isovector reaction\cite{kore2,kore4}}
The sum rules (10) and (11) correspond to the
moment at $n=0$. They are for the anti-symmetric combination 
under the interchange $a\leftrightarrow b$.  
Since the right-hand side is $Q^2$ independent, we obtain
\begin{equation}
 \int_{0}^{1}\frac{dx}{x}g_i^{[ab]}(x,Q^2)=\int_{0}^{1}\frac{dx}{x}g_i^{[ab]}(x,Q^2_0),\qquad
 for\quad i=1,2.
\end{equation}
Now, we take $Q_0^2=0$ and use the relation between the structure function
$G_1$ and the photo-production
\begin{equation}
 G_1^{ab}(\nu ,0)=-\frac{1}{8\pi^2\alpha_{em}}\{\sigma_{3/2}^{ab}(\nu ) - \sigma_{1/2}^{ab}(\nu )\}
=-\frac{1}{8\pi^2\alpha_{em}}\Delta\sigma^{ab}(\nu ).
\end{equation}
By setting $a=(1+i2)/\sqrt{2}, b=a^{\dagger}$, and separating out the
elastic contribution, we obtain the sum rule which relates the isovector part of
the $g_1$ and  the cross section in the
photo-production as in the Cabibbo-Radicati sum rule.\cite{CR}
Now the Regge theory predicts as 
$g_1^{[ab]}\sim \beta x^{-\alpha(0)}$ with $\alpha(0) \leq 0$,
and hence the sum rule is convergent.
However, the perturbative behavior like the DGLAP is
divergent. The double logarithmic $(log (1/x))^2$ resummation 
or the total resummation of $(log (1/x))^k$ gives the Regge
like behavior but the sum rule is also divergent.\cite{erm}
In such a situation, it is desirable to discuss the 
regularization of the sum rule and gives it a physical meaning even 
when the sum rule is divergent. 
Now, the regularization of the 
divergent sum rule has been known to be done by the analytical
continuation from the nonforward direction by assuming Regge type
behavior.\cite{dealwis} We first derive the finite sum rule in the small 
but sufficiently large $|t|$ region
by assuming the moving pole or cut. Then we subtract the singular pieces
which we meet as we go to the smaller $|t|$ from both hand sides of the
sum rule by obtaining the condition for the coefficient of the singular
piece. After taking out all singular pieces we take the limit $|t|\to
0$. Because of the kinematical structure in the course to derive the sum
rule, we can mimic this procedure in the forward direction by introducing the 
parameter which reflects the $t$ in the non-forward direction.
The sum rule obtained in this way can be transformed to the form
where the high energy behavior from both hand sides of the sum rule
is subtracted away.  Practically, if the cancellation at high energy is
effective, since the condition is needed only in the high energy limit, 
we consider that the sum rule holds irrespective of the condition. In this
way, we subtract the high energy behavior from both hand sides of the
sum rule as in the following way.
\begin{eqnarray}
\lefteqn{\int_0^1\frac{dx}{x}\{ g_1(x,Q^2) - f(x,Q^2)\} +
 \int_0^1\frac{dx}{x}f(x,Q^2)}&&\\\nonumber
&=& \int_0^1\frac{dx}{x}\{ g_1(x,Q_0^2) - f(x,Q_0^2)\} +
 \int_0^1\frac{dx}{x}f(x,Q_0^2),
\end{eqnarray}
where we set
\begin{equation}
f(x,Q^2) = \beta(Q^2)x^{-\alpha (0,\epsilon)} + f_1(x,Q^2)\qquad \alpha (0,\epsilon )=a - \epsilon.
\end{equation}
We take the limit $\epsilon \to a$ from the region above $a$.
\begin{equation}
 \int_0^1\frac{dx}{x}f(x,Q^2) = \frac{\beta (Q^2)}{\epsilon -a}
+\int_0^1\frac{dx}{x}f_1(x,Q^2).
\end{equation}
After taking out the pole term from both-hand side of the sum rule, we take $\epsilon \to 0$.
\begin{eqnarray}
\lefteqn{\int_0^1\frac{dx}{x}\{ g_1(x,Q^2) - f(x,Q^2)\}}&& \\\nonumber
&=& \int_0^1\frac{dx}{x}\{ g_1(x,Q_0^2) - f(x,Q_0^2)\} +
 \int_0^1\frac{dx}{x}\{f(x,Q_0^2) -f(x,Q^2)\}.
\end{eqnarray}
Now, for any $Q^2$, we take $f(x,Q^2)=g_1(x,Q^2)$ above $\nu_{c}^{Q} =m_pE_Q$ where
$E_Q=E_c+Q^2/2m_p$ and $f(x,Q^2)=0$ below it, where $E_Q$ is a cut-off
energy in the laboratory frame. We define
$x_c(Q^2)=\frac{Q^2}{2\nu_{c}^{Q}}$
and ,by setting $Q_0^2=0$ and separating out the Born term, we can
rewrite this relation for the proton target as
\begin{eqnarray}
\lefteqn{\int_{x_c(Q)}^{1}\frac{dx}{x}[2g_1^{1/2}(x,Q^2)-g_1^{3/2}(x,Q^2)]}&&\\\nonumber
&=&B(Q^2)-
\frac{m_p}{8\pi^2\alpha_{em}}\int_{E_0}^{E_c}dE[2\Delta\sigma^{1/2}-
\Delta\sigma^{3/2}]+K(E_c,Q^2),
\end{eqnarray}
\begin{equation}
B(Q^2)=\frac{1}{4}\{(\mu_p-\mu_n)-\frac{1}{1+Q^2/4m_p^2}G_{M}^{+}(Q^2)
[G_{E}^{+}(Q^2)+\frac{Q^2}{4m_p^2}G_{M}^{+}(Q^2)]\},
\end{equation}
\begin{equation}
G_E^+(Q^2)=G_E^p(Q^2)-G_E^n(Q^2), \qquad
G_M^+(Q^2)=G_M^p(Q^2)-G_M^n(Q^2),
\end{equation}
\begin{eqnarray}
K(E_c,Q^2)&=&-\int_{E_Q}^{\infty}\frac{dE}{E}[2g_1^{1/2}(x,Q^2)-g_1^{3/2}(x,Q^2)]\\\nonumber
&-&\frac{m_p}{8\pi^2\alpha_{em}}\int_{E_c}^{\infty}dE[2\Delta\sigma^{1/2}-\Delta\sigma^{3/2}],
\end{eqnarray}
\begin{equation}
 g_1^I,\Delta\sigma^I : {\rm isovector\; photon\; +\; proton\; \longrightarrow state\; with\; isospin\; I }.
\end{equation}
In case of the $g_2$, such a simple method to use the photo-reaction as the
regularization point can not be applied directly. 
Now in the relation
\begin{eqnarray}
\triangle\sigma^{ab}(\nu ,Q^2)&=&\sigma^{ab}_{3/2}(\nu ,Q^2) - \sigma^{ab}_{1/2}(\nu ,Q^2)\\\nonumber
&=&-\frac{8\pi^2\alpha_{em}}{K}\left(\frac{g_1^{ab}(x,Q^2)}{\nu} - 
\frac{m_N^2Q^2g_2^{ab}(x,Q^2)}{\nu^3}\right),
\end{eqnarray}
where $\displaystyle{K=(1 - \frac{Q^2}{2\nu})}$, if we differentiate it 
by $Q^2$ and take the limit $Q^2\to 0$, 
we obtain the relation
\begin{eqnarray}
\frac{g_2^{ab}(x,0)}{\nu} &=& \frac{g_1^{ab}(x,0)}{2m_N^2}
+\left. \frac{\nu}{m_N^2}\frac{\partial g_1^{ab}(x,Q^2)}{\partial Q^2}\right|_{Q^2=0}\\\nonumber
&+&\left.
  \frac{\nu^2}{8\pi^2m_N^2\alpha_{em}}\frac{\partial\triangle\sigma^{ab}(\nu, Q^2)}{\partial
Q^2}\right|_{Q^2=0}.
\end{eqnarray}
Thus we can relate $g_2^{ab}(x,0)/\nu$ to the experimentally measurable
quantity. Then, by setting $Q_0^2=0$,
we can rewrite the sum rule for the $g_2^{ab}$ by the same method 
as in the sum rule for the $g_1^{ab}$. 
In this way, we obtain the sum rule for the proton target as
\begin{equation}
\int_{x_c(Q)}^1\frac{dx}{x}g_2^{+-}(x,Q^2) =
B_2^{+-}(Q^2) +\int_{E_0}^{E_c}\frac{dE}{E}g_2^{+-}(x,0) +
K_2^{+-}(E_c,Q^2),
\end{equation}
where
\begin{equation}
g_2^{+-}(x,Q^2)=2g_2^{1/2}(x,Q^2)-g_2^{3/2}(x,Q^2).
\end{equation}
\begin{equation}
B_2^{+-}(Q^2) = \frac{Q^2}{16m_p^2}\frac{1}{1+\frac{Q^2}{4m_p^2}}
G_M^+(Q^2)(G_M^+(Q^2)-G_E^+(Q^2)).
\end{equation}
The same kind of the sum rules can be derived in the electroproduction,
if we extend the current commutation relation to the current
anti-commutation relation. This extension is possible as a stable hadron matrix element.
We explain this fact briefly in the following.
\section{DGS representation of the anti-commutator of the current}
Let us consider DGS representation\cite{DGS} by taking the scalar current.
\begin{eqnarray}
 C_{ab}(p\cdot q , q^2)& =&\int d^4x \exp (iqx)<p|[J_a(x) ,J_b(0)]|p>_c\\\nonumber
&=&\int d^4x\exp (iqx)\int_0^{\infty}d\lambda^2\int_{-1}^{1}d\beta h_{ab}
(\lambda^2,\beta )i\Delta (x,\lambda^2)\\\nonumber
&=&(2\pi )\int_0^{\infty}d\lambda^2\int_{-1}^{1}d\beta \delta ((q+\beta
p)^2-\lambda^2)\epsilon (q^0+\beta p^0)h_{ab}(\lambda^2,\beta ).
\end{eqnarray}
$C_{ab}$ can be written as
\begin{eqnarray}
 C_{ab}(p\cdot q , q^2)& =&\sum_n(2\pi )^4\delta^4(p+q-n)
\langle p|J_a(0)|n\rangle \langle n|J_b(0)|p\rangle \\\nonumber
&-&\sum_n(2\pi )^4\delta^4(p-q-n)
\langle p|J_b(0)|n\rangle \langle n|J_a(0)|p\rangle.
\end{eqnarray}
If we take the rest frame $p=(m,\vec{0})$,since
the first term is constrained as $m+q^0=n^0$, we obtain
$ q^0 \geqq M_s -m$, where $M_s$ is the lowest mass in the $s$ channel
continuum. Similarly, since
the second term is constrained as $m-q^0=n^0$ we obtain
$ q^0 \leqq m - M_u$, where $M_u$ is the lowest mass in the $u$ channel 
continuum. Hence if 
$m\leqq (M_s + M_u)/2$, the first term and the second term are disconnected.\\
\begin{minipage}{5cm}
\includegraphics[width=4.7cm]{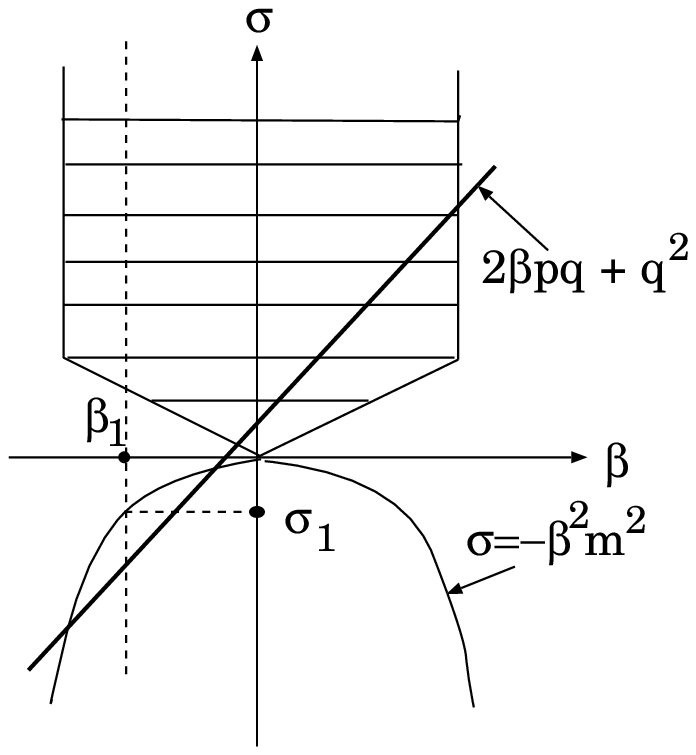}
\end{minipage}
\begin{minipage}{7cm}
Now, in the DGS representation (31), the support of the weight function
$h_{ab}(\lambda^2,\beta )$ lies in the shaded region in the figure. 
An integration path is $\sigma =2\beta p\cdot q +q^2$, where $\sigma = \lambda^2-\beta^2m^2$.
At the rest frame $ p\cdot q + \beta m^2=m(q^0+\beta p^0)$, hence
we obtain $\epsilon (q^0+\beta p^0)=\epsilon (p\cdot q +\beta m^2)$.
Since for the point $(\beta_1,\sigma_1)$ where 
$p\cdot q +\beta_1m^2=0$ and $\sigma_1=-\beta_1^2m^2$, the inequality
$\sigma_1 \geqq 2\beta_1p\cdot q + q^2$ holds,
the sign change always occurs in the causality forbidden region $\sigma
 < -\beta^2m^2$ as in the figure. In the $s$ channel, 
since $p\cdot q >0$, the slope is positive, hence
only the region $\epsilon (p\cdot q +\beta m^2)=1$ contributes.
Therefore $s$ channel and $u$ channel are disconnected.
Thus combined with the discussion after Eq.(32), we obtain
\end{minipage}
\begin{eqnarray}
(2\pi )\int_0^{\infty}d\lambda^2\int_{-1}^{1}d\beta \delta ((q+\beta
p)^2-\lambda^2)h_{ab}(\lambda^2,\beta )\theta (q^0+\beta p^0)\\\nonumber
=\sum_n(2\pi )^4\delta^4(p+q-n)
\langle p|J_a(0)|n\rangle \langle n|J_b(0)|p\rangle,
\end{eqnarray}
and
\begin{eqnarray}
(2\pi )\int_0^{\infty}d\lambda^2\int_{-1}^{1}d\beta \delta ((q+\beta
p)^2-\lambda^2)h_{ab}(\lambda^2,\beta )\theta (-(q^0+\beta p^0))\\\nonumber
=\sum_n(2\pi )^4\delta^4(p-q-n)
\langle p|J_b(0)|n\rangle \langle n|J_a(0)|p\rangle.
\end{eqnarray}
Eqs.(33) and (34) give us the DGS representation of the 
current anti-commutator of the current as
\begin{eqnarray}
 W_{ab}(p\cdot q , q^2)& =& \int d^4x \exp (iqx)<p|\{J_a(x) ,J_b(0)\}|p>_c\\\nonumber
&=&\int d^4x\exp (iqx)\int_0^{\infty}d\lambda^2\int_{-1}^{1}d\beta h_{ab}
(\lambda^2,\beta )i\Delta^{(1)} (x,\lambda^2)\\\nonumber
&=&(2\pi )\int_0^{\infty}d\lambda^2\int_{-1}^{1}d\beta \delta ((q+\beta
p)^2-\lambda^2)h_{ab}(\lambda^2,\beta ).
\end{eqnarray}
The application of this extension for the spin averaged quantity 
was given in \cite{kore_got}. For the spin dependent
part, it was given in \cite{kore} but the regularization of the sum rule
was not discussed. 
\section{Sum rules for the $g_1$ and the $g_2$ in the
 electroproduction \cite{kore3,kore4}}
Now, using the DGS representation for the current anti-commutator,
the (+ i) component of the anti-commutation relation of the
current on the null-plane was derived as
\begin{eqnarray}
<p,s|\{J_a^{+}(x),J_b^{i}(0)\}|p,s>_c|_{x^{+}=0,spin}\\\nonumber
=-\epsilon^{+i}_{\quad\alpha \beta}<p,s|\partial^{\alpha}[\Delta^{(1)}(x)
G_c^{5\beta}(x|0)]|p,s>_c,
\end{eqnarray}
where
\begin{equation}
 \Delta^{(1)}(x)|_{x^+=0}=-\frac{1}{2\pi}\ln |x^-|\delta
 (\vec{x}^{\bot}),
\end{equation}
and
\begin{equation}
 <p,s|S_c^{5\beta}(x|0)|p,s>_c=s^{\mu}S_c^5(p\cdot x,x^2)+p^{\mu}(x\cdot
 s)\bar{S}_c^5(p\cdot x,x^2)+x^{\mu}(x\cdot s)\tilde{S}_c^5(p\cdot x,x^2).
\end{equation}
By the same method as in Section 3, from Eq.(36) we can get sum rules for the
symmetric combination under the interchange $a\leftrightarrow b$,
hence for the electroproduction.
The sum rule of the $g_1$ for the proton target is
\begin{equation}
\int_{x_c}^1\frac{dx}{x}g_1^{ep}(x,Q^2) = B_1^{ep}(Q^2)
-\frac{1}{8\pi^2\alpha_{em}}\int_{\nu_0}^{\nu_c}d\nu\{
\sigma_{3/2}^{\gamma p} - \sigma_{1/2}^{\gamma p}\}+K_1^{ep}(E_c,Q^2),
\end{equation}
where
\begin{equation}
B_1^{ep}(Q^2)=\frac{1}{2}\{F_1^p(0)(F_1^p(0)+F_2^p(0))-F_1^p(Q^2)(F_1^p(Q^2)+F_2^p(Q^2))\},
\end{equation}
and
\begin{equation}
K_1^{ep}(E_c,Q^2)=\frac{1}{8\pi^2\alpha_{em}}\int_{\nu_c}^{\infty}d\nu\{
\sigma_{1/2}^{\gamma p} - \sigma_{3/2}^{\gamma p}\}
-\int_{\nu_c^Q}^{\infty}\frac{d\nu}{\nu}g_1^{p}(x,Q^2).
\end{equation}
In case of the $g_2$, we obtain
\begin{equation}
\int_{x_c(Q)}^1\frac{dx}{x}g_2^{ep}(x,Q^2))=
B_2^{ep}(Q^2)
+\int_{E_0}^{E_c}\frac{dE}{E}g_2^{ep}(x,0)
+K_2^{ep}(E_c,Q^2),
\end{equation}
\begin{equation}
B_2^{ep}(Q^2) =
 \frac{Q^2}{8m_p^2}\frac{1}{1+\frac{Q^2}{4m_p^2}}G_M^p(Q^2)(G_M^p(Q^2)-G_E^p(Q^2)).
\end{equation}
Then, we obtain
\begin{eqnarray}
\int_{x_c(Q)}^1\frac{dx}{x}(g_1^{ep}(x,Q^2) + g_2^{ep}(x,Q^2)) &=&
B_1^{ep}(Q^2)+B_2^{ep}(Q^2)\\\nonumber
+\int_{E_0}^{E_c}\frac{dE}{E}(g_1^{ep}(x,0)+g_2^{ep}(x,0))
&+& K_1^{ep}(E_c,Q^2)+K_2^{ep}(E_c,Q^2),
\end{eqnarray}
where
\begin{equation}
B_1^{ep}(Q^2)+ B_2^{ep}(Q^2) = \frac{1}{2}(\mu_p -G_M^p(Q^2)G_E^p(Q^2)).
\end{equation}
$K_1^{ep}(E_c,Q^2)$ in the sum rule (39) can be estimated with use of
parameters in Ref.\cite{Sim}.
We find that $K_1^{ep}(2,0.05)\sim -0.015$ and
$K_1^{ep}(2,0.1)\sim -0.029$. Therefore, as far as we consider the small
$Q^2$ region below $1$[GeV$^2$], this correction term
is very small and the sum rule should be satisfied by the contribution
below the energy $E_c=2$[GeV]. The contribution in this region
are the resonances, the nonresonant continuum, and the Born term. Further,
since the Born term changes very rapidly in this region, the sum of
the resonances and the nonresonant continuum
must also change rapidly. It is just in this region where the sign
change of the GDH sum was studied
experimentally.\cite{Fet}
A similar relation with the sum rule (39), but the quantity
corresponding to the moment at $n=1$
was given in Ref.\cite{ioffe}, because the GDH sum
rule\cite{Drell,Ger} and the Ellis-Jaffe\cite{Ellis} sum rule correspond to
the sum rule at $n=1$. The sum rule (39) is the exact relation 
corresponding to the moment at $n=0$, 
and the same kind of the sign change occur due to the
same physical origin as in the case of the moment at $n=1$, and hence
can be checked experimentally.  In such an analysis, 
if combined with the analysis of the
moment at $n=1$, the extraction of the $g_1$
from the experimental data of the assymetry in the resonance region 
at small $Q^2$ will become very important. 
Now, let us consider another model of the $g_1$ in Ref.\cite{ST} from
our sum rules (39), (42),(44).
The magnitude of the Born term contributions in the moment at $n=0$ for the
$g_1^{ep}$ and the $(g_1^{ep}+g_2^{ep})$ are very similar,
but that of the $g_2^{ep}$ is very small compared with these
since it is proportional to $Q^2$. However, if 
this Born term is divided by $Q^2/2$, it has a
finite limit as $Q^2\to 0$, and has an interesting behavior. 
This quantity is the one which appears in the
Schwinger sum rule for the $g_2^{ep}$ given as\cite{Schw,DJT,BC}
\begin{equation}
\frac{-1}{4m_p^2+Q^2}G_M^p(Q^2)(G_M^p(Q^2)-G_E^p(Q^2))
+\int_{\nu_0(Q)}^{\infty}d\nu G_2^{ep}(\nu ,Q^2) = 0,
\end{equation}
where we separate the Born term in this sum rule. It should be
noted that this sum rule is nothing but the sum rule (12)
derived from the $+i$ component in the current commutation relation.
At large $Q^2$, because the Born term becomes negligible, 
we have the relation for the inelastic part.
\begin{equation}
I(Q^2)=\int_{\nu_0(Q)}^{\infty}d\nu G_2^{ep}(\nu ,Q^2)=\frac{2}{Q^2}
\int_{0}^{1}dxg_2^{ep}(x,Q^2)=0.
\end{equation}
Thus we can consider the main contribution in the continuum
part in the Schwinger sum rule comes from a relatively low
energy region. Therefore, in the sum rule given as
\begin{equation}
\int_{\nu_0(Q)}^{\infty}d\nu G_2^{ep}(\nu ,Q^2) -
 \int_{\nu_0}^{\infty}d\nu G_2^{ep}(\nu ,0)
= B_S^{ep}(Q^2),
\end{equation}
where 
\begin{equation}
B_S^{ep}(Q^2) = \frac{1}{4m_p^2+Q^2}G_M^p(Q^2)(G_M^p(Q^2)-G_E^p(Q^2))
-\frac{\mu_p(\mu_p-1)}{4m_p^2}
\end{equation}
the main contribution on the left hand side comes from the
low $Q^2$ region. Since the Born term contribution $B_S(Q^2)$ changes
rapidly in this region, the left hand side of the sum rule also
changes rapidly. Since we have the relation $\nu = Q^2/2$ at the elastic
point,  $B_S^{ep}(Q^2)$ is related to $B_2^{ep}(Q^2)$ as
\begin{equation}
B_S^{ep}(Q^2) = \frac{2}{Q^2}B_2^{ep}(Q^2) - \left. \left\{
\frac{2}{Q^2}B_2^{ep}(Q^2)\right\}\right|_{Q^2=0}
\end{equation}
Now the contribution to the quantity
\begin{equation}
\int_{x_c(Q)}^1\frac{dx}{x}g_2^{ep}(x,Q^2) - 
  \int_{x_c}^1\frac{dx}{x}g_2^{ep}(x,0)
\end{equation}
in the sum rule for the $g_2^{ep}$ corresponding to 
the moment at $n=0$ comes from the low energy region
and we can expect it roughly given by $B_2^{ep}(Q^2)$. 
Thus this sum rule and the Schwinger sum rule gives
us the same picture that the rapid behavior of the elastic
is compensated by the rapid behavior of the resonance and the
nonresonant continuum.
Now if we plot the Born term contributions $B_1^{ep}(Q^2),B_1^{ep}+B_2^{ep}(Q^2),$
and $-B_S^{ep}(Q^2)$, at small $Q^2$ below 1[GeV$^2$] 
we find that these three functions behave very
similarly. The difference between $B_1^{ep}(Q^2)$
and $-B_S^{ep}(Q^2)$ is very small and moreover the difference is almost
constant.\\
Though the moments which give $B_S^{ep}(Q^2)$ and $B_1^{ep}(Q^2)$
are different, we see that the behavior of the integral of
$\{-2g_2^{ep}(x,Q^2)/Q^2+(2g_2^{ep}(x,Q^2)/Q^2)|_{Q^2=0}\}$ and that of
$\{g_1^{ep}(x,Q^2)/x - (g_1^{ep}(x,Q^2)/x)|_{Q^2=0}\}$ in the small $Q^2$
region is very similar. Since the latter is related to the sign change
of the GDH sum, this fact may suggest that
the $g_2^{ep}$ is related to this phenomena. However, in our approach,
we have no direct relation between the $g_1^{ep}$ and the $g_2^{ep}$.
\section{Summary}
Sum rules of the spin dependent structure functions 
based on the canonical quantization on the null-plane
corresponding to the moment at $n=0$ are regularized. 
By taking the regularization point at $Q^2=0$ and relating to the
photoproduction, spin dependent structure functions $g_1$ and $g_2$ 
at small $Q^2$ in the low and the intermediate energy region
are shown to be severely constrained. For example,    
we find, in the small $Q^2$ region near $Q^2\sim 0.1$(GeV/c)$^2$,
that the integral $\displaystyle{\int_{x_c}^1\frac{dx}{x}g_1^{p}(x,Q^2)}$
becomes zero and that it
changes the sign from the negative to the positive. 
This behavior is caused by the 
rapid change of the resonances to compensate
the rapid change of the elastic to satisfy the sum rule. 
It is this rapid change of the resonances which gives 
the sign change of the GDH sum. 
In addition to this tight connection among the resonances,
the elastic and the nonresonant continuum, we also find that this behavior is very
similar between the $g_1^{ep}$ and the $g_2^{ep}$ as pointed out in Ref.\cite{ST}.

\end{document}